\documentclass[aps,showpacs,twocolumn,superscriptaddress,nofootinbib,floatfix]{revtex4-1}

\usepackage{graphicx}
\usepackage{amsfonts}
\usepackage{amsmath,amssymb}
\usepackage{bm}
\usepackage{color}
\usepackage{float}
\usepackage{dcolumn}
\usepackage{cancel}

\begin{document}

\title{Information content of the weak-charge form factor}

\author{P.-G. Reinhard}
\affiliation{Institut f\"ur Theoretische Physik II, Universit\"at Erlangen-N\"urnberg,  
Staudtstrasse 7, D-91058 Erlangen, Germany}
\author{J. Piekarewicz}
\affiliation{Department of Physics, Florida State University,
                  Tallahassee, Florida 32306, USA}
\author{W. Nazarewicz}
\affiliation{Department of Physics \&
Astronomy, University of Tennessee, Knoxville, Tennessee 37996, USA}
\affiliation{Physics Division, Oak Ridge National Laboratory, Oak Ridge, Tennessee 37831, USA}
\affiliation{Faculty of Physics, University of Warsaw, ul. Ho\.za 69, 00-681 Warsaw, Poland}
\author{B. K. Agrawal}
\affiliation{Saha Institute of Nuclear Physics, Kolkata 700064, India}
\author{N. Paar}
\affiliation{Physics Department, Faculty of Science, University of Zagreb, Zagreb, Croatia}
\author{X. Roca-Maza}
\affiliation{Dipartimento di Fisica, Universit\`a degli Studi di Milano \
and INFN, 20133 Milano, Italy}

\date{29. July 2013}

\begin{abstract}
\begin{description}
\item[Background]
Parity-violating electron scattering provides a model-independent determination of the nuclear weak-charge form factor that has widespread implications across such diverse areas as fundamental symmetries, nuclear structure, heavy-ion collisions, and neutron-star structure. 
\item[Purpose] 
We assess the impact of precise measurements of the weak-charge form factor of  ${}^{48}$Ca and ${}^{208}$Pb on a variety of nuclear observables, such as the neutron skin and the electric-dipole polarizability. 
\item[Methods]
We use the nuclear Density Functional Theory with several accurately calibrated non-relativistic and relativistic energy density functionals. To assess the degree of correlation between nuclear observables and to explore systematic and statistical uncertainties on theoretical predictions, we employ the chi-square statistical covariance technique.
\item[Results] 
We find a strong correlation between the weak-charge form factor and the neutron radius, that allows for an accurate determination of the 
neutron skin of neutron-rich nuclei. We determine the optimal range of the momentum transfer $q$ that maximizes the information content of the measured weak-charge form factor  and quantify the uncertainties associated with the strange quark contribution. Moreover, 
we  confirm the role of the electric-dipole polarizability as a strong isovector indicator. 
\item[Conclusions]
Accurate measurements of the weak-charge form factor of ${}^{48}$Ca and ${}^{208}$Pb will have a profound impact on many aspects of nuclear theory and hadronic measurements of neutron skins of exotic nuclei at radioactive-beam facilities.
\end{description}
\end{abstract}

\maketitle

\section{Introduction}\label{intro}

The Lead Radius EXperiment (``PREX'') at the Jefferson Laboratory has
used parity-violating electron scattering to probe the weak-charge
density distribution in
$^{208}$Pb\,\cite{Abrahamyan:2012gp,Horowitz:2012tj}. Given that the
weak charge of the neutron is much larger than that of the proton,
parity-violating electron scattering provides a clean probe of neutron
densities\,\cite{Donnelly:1989qs}. The parity-violating asymmetry
calculated within Born approximation, although qualitative, is
directly proportional to the weak-charge form factor, which in turn is
obtained from the weak-charge density by a Fourier transform. This
direct relation is preserved 
in  calculations that account for Coulomb
distortions \cite{(Hor01),roca-maza11}.  PREX
measured the weak-charge form factor of $^{208}$Pb at a momentum
transfer of $q_{{}_{\rm PREX}}\!=\!0.475\,{\rm fm}^{-1}$ to
be\,\cite{Horowitz:2012tj}
\begin{equation}
 F_{W}(q_{{}_{\rm PREX}}) = 0.204 \pm 0.028 \,.
 \label{FwPREX}
\end{equation}
By making some  assumptions pertaining to the form factor, PREX was able to provide the first 
determination of the neutron-skin  of 
${}^{208}$Pb\,\cite{Abrahamyan:2012gp}:
\begin{equation}
r_\mathrm{skin}^{208}\!=\!r_{n}^{208}\!-\!r_{p}^{208}\!=\!{0.33}^{+0.16}_{-0.18}\,{\rm fm},
 \label{NSkin}
\end{equation}
where $r_{n}^{208}(r_{p}^{208})$ is the neutron (proton)
root-mean-square (rms) radius of ${}^{208}$Pb. Although PREX demonstrated
excellent control of systematic errors, the statistical accuracy of
the measurement was compromised. Fortunately, the PREX collaboration
has made a successful proposal (``PREX-II")\,\cite{PREXII} that will allow them to
reach their original goal of 0.06\,fm in the experimental
uncertainty.  Given
that PREX demonstrated that model-independent measurements of the
weak-charge form factor in heavy nuclei are now feasible, it is
pertinent to ask whether a measurement in a different neutron-rich
nucleus could prove advantageous. Indeed, the case of ${}^{48}$Ca
seems particularly attractive for several reasons. First, ${}^{48}$Ca
is a doubly-magic nucleus that is already within the reach of
ab-initio calculations\,\cite{(Hol12),Hagen:2012fb}. Thus, the recently approved Calcium Radius
EXperiment (``CREX'')\,\cite{CREX} could provide a critical bridge between
ab-initio approaches and density-functional theory. Second, by
providing this kind of bridge, CREX will help elucidate the character
of the three-nucleon force, or the density dependence of the energy density functional, which play a critical role in determining
the limits of the nuclear landscape \cite{(Erl12),(Erl13),(Eks13),(Wie13)} and properties of nuclear and neutron matter \cite{(Ste12),(Tew13),(Heb13)}.  Finally, CREX -- together with PREX-II --
will provide calibrated benchmarks for hadronic measurements of neutron
skins at radioactive beam facilities.  Note that the CREX
collaboration has made a successful proposal
to measure the neutron radius of ${}^{48}$Ca using parity-violating electron 
scattering with an unprecedented accuracy of 0.02\,fm\,\cite{CREX}.
This has a great potential to guide further theoretical developments
\cite{(Pie12),(Kor13a)}.

Among the great variety of nuclear structure models, self-consistent
mean-field (SCMF) models  rooted in the nuclear Density Functional Theory (DFT) provide the best compromise between accuracy,
computational expediency, and universality (by covering the greatest
range of accessible nuclei)\,\cite{(Ben03),Vre05aR}. This
survey is concerned with the performance of SCMF models in relation to the
weak-charge form factor and its possible impact on improving 
energy density functionals (EDF)
that  are
at the heart of the SCMF approaches.  The structure of all EDFs can be motivated on formal grounds
by invoking methods such as density-matrix expansion that may be
tested against ab-initio approaches. Although enormous progress has
been made in developing ab-initio techniques, at present they are of
limited use in building the spectroscopic quality of EDFs. 
Thus, the coupling constants of EDFs, i.e., model parameters,  must
be determined empirically through a fit to selected nuclear data.
Once determined, often through the minimization of an appropriate
objective function, the parameters are universal in that the same
functional can in principle be applied to all nuclei, nuclear reactions, and neutron stars. Such an empirical fit
also  provides valuable information on the statistical uncertainties of the model
parameters and the correlations between them. This is of
particular importance  in the context  of the isovector sector of the
EDFs. Indeed, whereas the isoscalar sector of the density functional is fairly well
determined by the pool of available nuclear data (such as ground-state masses and
charge radii), the isovector sector is hindered by the sparsity of
high-quality data that are sensitive to the neutron-proton
asymmetry. This implies that all isovector-sensitive observables, such
as the neutron skin and electric-dipole polarizability of
neutron-rich nuclei, are predicted with large theoretical
uncertainties. However, those uncertainties can be turned into an
advantage by allowing us to explore correlations between different
observables. Indeed, it was through such a statistical 
covariance analysis that a strong correlation
between the electric dipole polarizability and the neutron skin 
of ${}^{208}$Pb has been established\,\cite{(Rei10)}. We wish to emphasize  that theoretical
uncertainties and correlations among observables are estimated within
a given model by computing the covariance matrix associated with the
minimization of the associated objective  function\,\cite{(Rei10),(Klu09),(Kor10),*(Kor12),(Fat11),(Gao13),(Kor13a)}.
Although such an approach provides the  statistical
uncertainties and correlations, it cannot assess the systematic errors
that reflect constraints and limitations of a given model. Such
systematic uncertainties can only emerge by comparing different
models\,\cite{(Pie12),(Kor13a),(Naz13)}. It is the aim of the present paper to apply
both statistical and systematic (or trend) analyzes to investigate uncertainties
and correlations associated to the weak-charge form factor at the
momentum transfers of relevance to PREX-II and CREX.

The manuscript has been organized as follows. In Sec.\,\ref{sec:formalism}
we develop the formalism required to carry out the correlation analysis for
a variety of accurately-calibrated SCMF models. Results are presented
in Sec.\,\ref{sec:results} for the correlations between various observables
using both a trend and a covariance analysis. We summarize our results
and present the outlook for the future in Sec.\,\ref{sec:conclusions}.

\section{Formalism}
\label{sec:formalism}

In this section we introduce the SCMF models  used
in the present survey, the observables  discussed, and the
details of the statistical covariance analysis.

\subsection{SCMF models}

In this survey we compare results from two different nuclear
EDFs: the non-relativistic Skyrme-Hartree-Fock (SHF) and 
the relativistic mean-field (RMF) models. There is some variety 
among the relativistic models. Here we will consider the standard 
non-linear (NL) RMF, its point-coupling (PC) variant, its extension 
from Florida State University (FSU), and finally the relativistic
model with density dependent meson-nucleon couplings (DDME).
We now briefly summarize the essential features and fitting 
protocols of the various functionals.

The SHF model uses an  EDF which is constructed from
baryon, spin-orbit, and kinetic-energy
densities. Each interaction term (including the spin-orbit term) appears in
isoscalar and isovector form.  This provides the model with great
flexibility in the isovector channel. Note that SHF has, unlike the
RMF model, explicit independent parameters for the spin-orbit coupling.  The
 functional has altogether about ten free
parameters. Pairing is modeled by a density dependent contact
interaction having three further free parameters. For details see
\cite{(Ben03),(Klu09),(Kor10)}. Here, we shall use  the SV parameterizations
from Ref.\,\cite{(Klu09)}. These fits were done to a large pool of
semi-magic nuclei that were checked to have negligible correlation
effects \cite{(Klu08)}. The observables included in the SV optimization database are: binding energy
(70 entries), rms charge radius (50 entries),
charge diffraction radius (28 entries), charge surface thickness (26
entries), neutron and proton pairing gaps (37 entries), and spin-orbit
splittings of single-particle energies (7 entries).  An objective function
$\chi^2$ was calibrated to these data and its minimization yields the
SV-min parametrization that will be used here for the correlation
analysis (see below). We have also provided a couple of further
parametrization with systematic variation of nuclear matter properties
(incompressibility $K$, isoscalar effective mass $m^*/m$, symmetry
energy at the saturation density $J$ and TRK sum rule enhancement
$\kappa$ related to isovector effective mass). 
To this end, these four
properties are constrained in a fit using the same data as for
SV-min. Several fits of that sort are run producing four chains each
one varying exclusively one of these nuclear matter properties. This
set of parametrizations is used for the trend analysis.
 
The RMF model consists of Dirac nucleons interacting via the exchange of 
three {\it``mesons"}: an isoscalar-scalar $\sigma$-meson, an isoscalar-vector 
$\omega$-meson, and an isovector-vector $\rho$-meson. The corresponding
baryon densities become the sources for the meson-field equations that are
solved at the mean-field level. In turn, the meson fields provide the scalar 
and vector potentials that enter into the Dirac equation. This procedure is
repeated until self-consistency is achieved\,\cite{Serot:1984ey}. A quantitatively 
successful RMF model emerges when these three Yukawa couplings were 
augmented by a non-linear (NL) self-coupling of the $\sigma$ 
meson\,\cite{Boguta:1977xi}. This had led to some successful applications 
throughout the nuclear chart, first NL1\,\cite{Rei86a} and then  
NL3\,\cite{Lalazissis:1996rd}; for some reviews see Refs.\,\cite{Rei89aR,Rin96aR}. 
However, with increasing demands on quality predictions,  several deficiencies of the original non-linear 
models became apparent. For example, both the incompressibility
of symmetric nuclear matter and the slope of the symmetry energy were notoriously 
high as compared to SHF models. In particular, this hindered the description 
of giant monopole resonances (GMR)  over a larger mass 
range\,\cite{Piekarewicz:2003br}. In particular, NL3 overestimates the location 
of the breathing mode in ${}^{90}$Zr -- a nucleus with a well-developed GMR peak 
but  small neutron-proton asymmetry.

In response to these shortcomings, a new FSUGold parametrization was developed\,\cite{Todd-Rutel:2005fa} by extending the  NL3 model
by two additional terms in order to soften both the
equation of state of symmetric nuclear matter and the symmetry energy.
Following standard practices, FSUGold was accurately calibrated through 
the  minimization of  $\chi^2$ constrained by the binding 
energies and charge radii of magic nuclei, as well as some bulk 
properties of nuclear matter. The slight extension 
of the model allowed FSUGold to generate a smaller incompressibility coefficient
and a softer symmetry energy, which proved essential in reproducing simultaneously 
the GMR in both ${}^{90}$Zr and ${}^{208}$Pb, as well as the isovector giant 
dipole resonance in ${}^{208}$Pb. We note, however, that at the time of the calibration
of the FSUGold interaction, no covariance matrix was extracted. Hence, the
correlation coefficients predicted in the present work by FSUGold (or ``FSU" for short)
are obtained from the simplified covariance analysis presented in
Ref.\,\cite{(Fat11)}. As in the non-relativistic case, trends will
also be studied by producing NL and FSU chains (or ``families")
through a systematic variation of the model parameters. In the case of
NL and FSU, the systematic variations are implemented by only varying
the two isovector parameters of the model; the isoscalar sector
remains intact. For details on the implementation we refer the reader
to Refs.\,\cite{Horowitz:2000xj,Horowitz:2001ya}.

We also employ two other variants of the RMF model. The RMF-DDME
functional is based on the standard form of Yukawa-coupled
nucleon-meson interactions, but with the coupling constants
supplemented with an elaborate density
dependence\,\cite{Nik02}. Modeling the density dependence introduces
four additional free parameters which brings to 8 the total number of
parameters in RMF-DDME. In the RMF-PC model one effectively eliminates
the mesons by making their masses much larger than any scale in the
problem. In this model, nucleons interact via four-fermion contact
interactions or equivalently, via point coupling terms that are
quadratic in the various baryon densities\,\cite{Bue02}.  Similar to
the NL models, non-linearities are introduced through cubic and
quartic terms in the scalar density. Finally, to compensate for the
finite range of the (missing) meson fields, the model is supplemented
with derivative (or gradient) coupling terms involving the two vector
densities. This amounts to 9 free parameters for the RMF-PC model.
Both in DDME and PC variants of the RMF model, the pairing force is
introduced by using the BCS approximation with empirical pairing gaps.
Finally, in both cases, optimal parametrizations were obtained by
fitting to the dataset that includes ground-state binding energies,
charge radii, diffraction radii, and surface thickness of 17 spherical
nuclei ranging from $^{16}$O to $^{214}$Pb.

\subsection{Nuclear observables}
  
As we have seen above, basic ground-state observables, such as binding
energies and charge radii are critical inputs for the calibration of
the functional.  Here we 
consider additional observables that were not included in the
calibration and whose correlations we wish to explore. These are:
(i) the root-mean-square neutron radius $r_n$; (ii) the
neutron skin  $r_\mathrm{skin}$; (iii) the weak-charge form factor $F_{W}(q)$; and
(iv) the electric dipole polarizability $\alpha_{D}^{\mbox{}}$. The rms radii are
computed from the $r^2$ weighted density distribution. The weak-charge
form factor $F_{W}(q)$ is obtained from the Fourier transform of the
corresponding weak-charge density  with the calibration 
 $F_{W}(0)=1$ (for details see
Appendix\,\ref{sec:FW}). Finally, the electric dipole polarizability is
computed in a random-phase approximation (RPA) from the inverse-energy
weighted dipole strength
$\alpha_D=2\sum_{n}(|\langle\Phi_n|\hat{D}|\Phi_0\rangle|^2/E_n)$,
where $n$ runs over all excited states of the system. We note that the
RPA is the consistent linear response of the mean-field ground state
to external perturbations. We will compute these observables with a
variety of EDFs for $^{48}$Ca and
$^{208}$Pb in an effort to emphasize the importance of CREX and
PREX-II.
Besides these observables in finite nuclei, we will 
consider key response properties of symmetric nuclear matter: the
incompressibility $K$, the symmetry energy $J$, and the density
dependence of the symmetry energy $L$.
 
\subsection{Correlating observables}

Each EDF is
characterized by about a dozen free parameters
$\mathbf{p}=(p_1,\ldots,p_F)$ that are calibrated to a host of
observables from finite nuclei. The most efficient and systematic
implementation of the calibration procedure is through a least-squares
fit. The fitting procedure starts with the definition of an objective function
(quality measure)  $\chi^2(\mathbf{p})$ that is computed by accumulating the
squared residuals of calculated observables relative to
the experimental data 
\begin{equation}
\chi^2 = \sum_{i=1}^{N}\left[\frac{\mathcal{O}_i^{\rm exp.}-\mathcal{O}_i^{\rm theo.}(\mathbf{p})}{\sigma_i}\right]^2,
\label{chi2}
\end{equation}
and weighted by the corresponding one standard deviation $\sigma_i$
associated with the $i-$th observable.
The optimum parametrization $\mathbf{p}_0$ is the
one that minimizes $\chi^2$ with the minimum value given by
$\chi^2_0\equiv\chi^2(\mathbf{p}_0)$.  Model parameters $\mathbf{p}$
which lie in the immediate vicinity of $\mathbf{p}_0$ also provide a
good description of the experimental data.  Moreover, the trends in
this vicinity encode a wealth of useful information which we can
exploit in a covariance analysis. Specifically, the range of
``reasonable'' parametrizations is defined to cover all
model-parameters $\mathbf{p}$ for which
$\chi^2(\mathbf{p})\leq\chi^2_0\!+\!1$ \cite{(Bra07aB)}. Given that such range of
parameters is usually rather small, we can expand $\chi^2$ in a power
series around $\mathbf{p}_0$. That is, up to second order in
$(\mathbf{p}\!-\!\mathbf{p}_0)$ we obtain
\begin{equation}\label{chi2a}
  \chi^2(\mathbf{p})\!-\!\chi^2_\mathrm{0}
  \approx \sum_{i,j=1}^F 
(\mathbf{p}\!-\!\mathbf{p}_0)_{i}\,\mathcal{M}_{ij}(\mathbf{p}\!-\!\mathbf{p}_0)_{j}\,,
\end{equation}  
where $\mathcal{M}_{ij}$ is the matrix of second derivatives:
\begin{equation}\label{Mij}
 \mathcal{M}_{ij}=\frac{1}{2}\partial_{p_i}\partial_{p_j}\chi^2(\mathbf{p})\Big|_{\mathbf{p}_0}\,.
\end{equation}
The reasonable parametrizations thus fill the confidence ellipsoid given by  
(see Sec.~9.8 of \cite{(Bra07aB)})
\begin{equation}\label{confidence}
  (\mathbf{p}\!-\!\mathbf{p}_0)^{T}\hat{\mathcal{M}}(\mathbf{p}\!-\!\mathbf{p}_0)
  \leq 1\,.
\end{equation}
It is now interesting to examine the impact of the formulation on
physical observables. Each set of model parameters $\mathbf{p}$
determines the functional and thus any observable $A$ predicted by
such functional may be considered a function of the parameters, {\sl
  i.e.,} $A\!=\!A(\mathbf{p})$. For a Gaussian distribution
$\exp{\left[-\chi^2(\mathbf{p})\right]}$ of the different
parametrizations $\mathbf{p}$ around the minimum $\mathbf{p}_0$, the
central value of the observable is given by
$A_{0}\!\equiv\!A(\mathbf{p_{0}})$ and there is an uncertainty in the
value of $A$ as one varies the $\mathbf{p}$ within the confidence
ellipsoid.  We now assume for simplicity that the observable varies
slowly with $\mathbf{p}$ within the relevant range, so that we can
estimate its uncertainty through a linear estimate. That is,
\begin{equation}\label{DeltaA}
 A(\mathbf{p}) \approx A_{0}+(\mathbf{p}\!-\!\mathbf{p}_0)\cdot
 \partial_\mathbf{p}A\Big|_{\mathbf{p}_0}\,. 
\end{equation}
The Gaussian-weighted average over the
parameter landscape yields the combined uncertainties of two 
observables $A$ and $B$, i.e., their covariance:
\begin{equation}\label{cova}
  \overline{\Delta A\,\Delta B} =
  \sum_{ij}\Big(\partial_{p_i}A\Big)(\hat{\mathcal{M}}^{-1})_{ij}
               \Big(\partial_{p_j}B\Big)\,.
\end{equation}
In the case of $A$=$B$, then Eq.~(\ref{cova}) gives the variance
$\overline{(\Delta A)^{2}}$, which defines the uncertainty of $A$. Variance 
and covariance are useful concepts that allow to estimate the impact of 
an observable on the model and its fit. We will exploit this in two ways by means of
(i) a trend analysis and (ii) a covariance analysis. In the trend analysis, 
parameters of the optimum model are modified according to a change
in a given bulk parameter of infinite nuclear matter and then  
the response of the observable of interest is monitored. For example, one could fix 
the slope of the symmetry energy $L$, then constrain the remaining model parameters
to reproduce this value, and finally monitor how the neutron skin
 of ${}^{208}$Pb responds to this change. Such a strategy helps elucidate 
systematic differences among the predictions of the models. In
the covariance (or correlation) analysis, on the other hand, only information from the 
properly extracted covariance matrix $\hat{\mathcal{M}}^{-1}$ is used
to compute statistical correlations within a given optimum model. 
A useful dimensionless statistical measure of correlation between two
observables is the Pearson product-moment correlation 
coefficient\,\cite{(Bra07aB)}:
\begin{equation}
  {c}_{AB} =
  \frac{|\overline{\Delta A\,\Delta B}|}
       {\sqrt{\overline{\Delta A^2}\;\overline{\Delta B^2}}}\,.
\label{eq:correlator}
\end{equation}
In particular, a value ${c}_{AB}\!=\!1$ means that the two observables are fully 
correlated whereas a value of ${c}_{AB}\!=\!0$ means that they are
uncorrelated. Note that we do not distinguish between perfect correlation 
${c}_{AB}\!=\!+1$ and perfect anti-correlation ${c}_{AB}\!=\!-\!1$.

\section{Results}
\label{sec:results}

In this section we present results for the correlations between
observables.  We start with the conceptually simpler trend analysis
and continue with the more quantitative covariance analysis.

\subsection{Trend analysis}

\begin{figure}[ht]
\begin{center}
\includegraphics[width=0.99\linewidth]{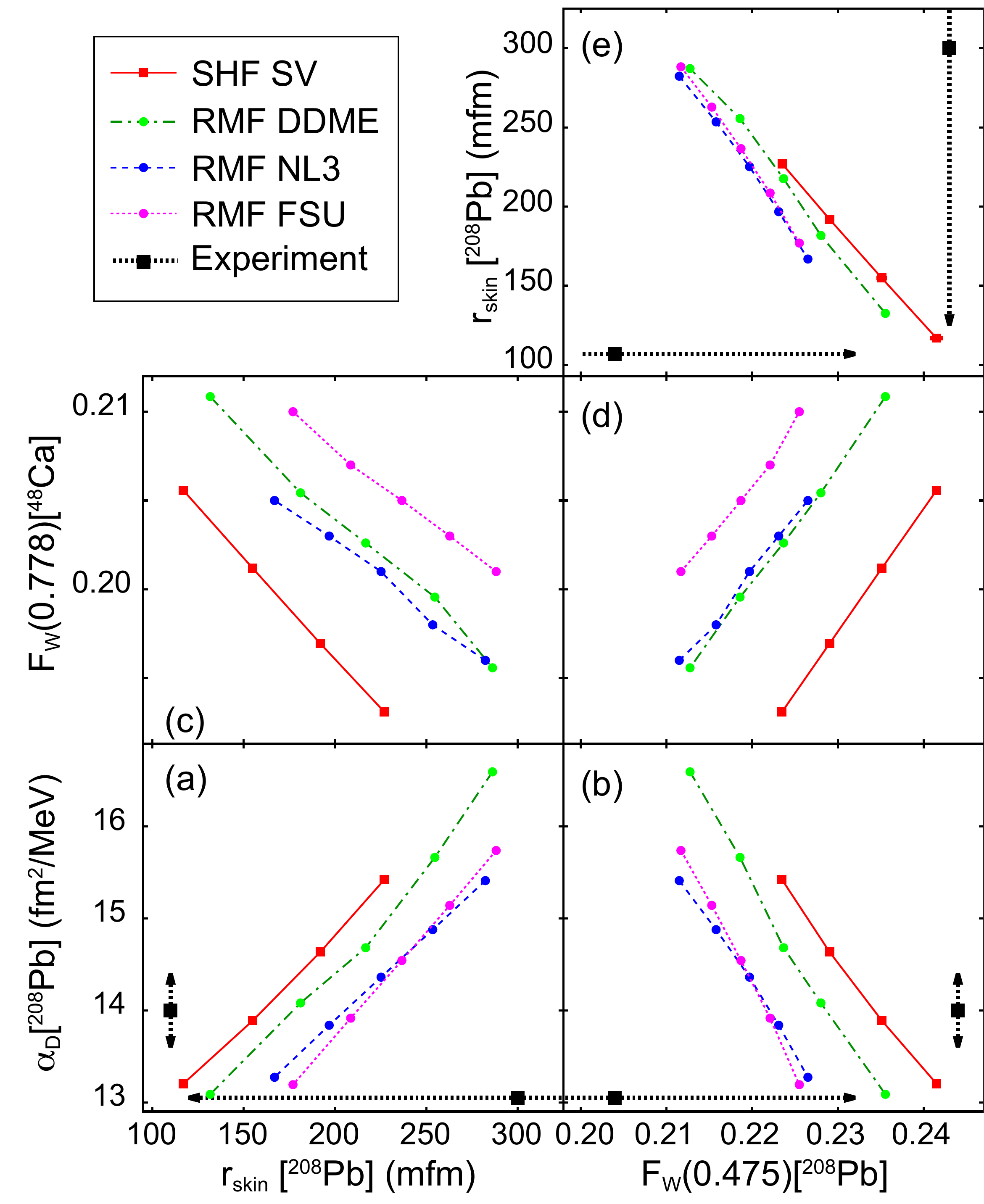}
\hspace*{1em}
 \caption{(Color online) Systematic trends as predicted by the various families of models for
 the following isovector observables:
 the neutron skin, electric-dipole polarizability,  weak-charge form factor
 $F_W(q\!=\!0.475\,{\rm fm}^{-1})$ of ${}^{208}$Pb, and the weak-charge form factor
 $F_W(q\!=\!0.778\,{\rm fm}^{-1})$ of ${}^{48}$Ca. Experimental values are indicated by black 
 squares with error bars ($r_\mathrm{skin}^{208}$ and $F_{W}^{208}$ from Refs.\,\cite{Abrahamyan:2012gp,Horowitz:2012tj} and 
  $\alpha_{D}^{208}$ from Ref.\,\cite{(Tam11)}).}
\label{Fig1}
\end{center}
\end{figure}

A simple way to visualize the mutual dependence between isovector
observables is to produce sets of parametrizations with systematically
varied symmetry energy $J$ and to study the behavior of a pair
of observables along those sets (see, e.g., Refs.~\cite{(Klu09),(Naz13)}). 
In the following, we compare  the predictions of four different families of 
models (SHF, DDME, NL3, FSU) that cover a systematic variation of $J$. 
As displayed in Fig.\,\ref{Fig1}, such a variation is reflected in
systematic changes to strong isovector indicators, such as 
$r_{\rm skin}^{208} \equiv r_{\rm skin}$[$^{208}$Pb]
 and the associated weak-charge
form factor $F_{W}^{208}\!\equiv\!F_W(q_{{}_{\rm PREX}})$[$^{208}$Pb]. 
Very strong correlations appear for {\it all} pairs of observables and for {\it all} model families.

As alluded earlier, $F_{W}^{208}$ is strongly sensitive to the density dependence
of the symmetry energy and this is reflected in its strong correlation
with $r_{\rm skin}^{208}$, as displayed in Fig.\,\ref{Fig1}(e).
This strong correlation appears universal as {\it all models}
lie practically on one line, suggesting that the experimentally
extracted weak-charge from factor 
$F_{W}^{208}$ from the parity violating 
asymmetry provides a strong constraint on $r_\mathrm{skin}^{208}$
\cite{Abrahamyan:2012gp,roca-maza11,Furnstahl:2001un,Horowitz:2012tj}.  
However, we observe a weaker inter-model correlation
between these two observables and the electric dipole polarizability
$\alpha_{D}$[$^{208}$Pb], see panels (a) and (b), and the
weak-charge form factor of $^{48}$Ca ($F_{W}^{48}$) in  Fig.\,\ref{Fig1}(c)
and (d). The correlation between $\alpha_{D}^{208}$ and
$r_\mathrm{skin}^{208}$ has been studied in Ref.\,\cite{(Pie12)} that has 
confirmed $\alpha_{D}^{208}$ as a key isovector indicator (see also 
discussion in Refs.~\cite{(Rei13),roca-maza13}).
We note that the electric dipole polarizability in ${}^{208}$Pb 
was recently measured at the Research Center for Nuclear Physics 
(RCNP) using polarized proton inelastic scattering at forward angles. 
The reported value from such a landmark experiment is\,\cite{(Tam11)}
\begin{equation}
  e^2\alpha_{D}^{208} = (20.1 \pm 0.6)\, {\rm fm}^{3} \;.
\label{DipPol208}
\end{equation}
Note that there two conventions for the units of the
polarizability are commonly used. Some references \cite{(Tam11),(Pie12)}, augment it with
the charge factor $e^2=1.44$\,MeV\,fm, thus expressing  $\alpha_{D}$
in units of fm$^3$. Others, e.g., \cite{(Rei10)} and this work, use the
dipole operator without charge, which leads to units {fm}$^2$/{MeV}.

In this work, we look  for the first time into  the
correlation between the weak-charge form factors of $^{48}$Ca and
$^{208}$Pb displayed in Fig.\,\ref{Fig1}(d). In particular, we note a significant model variance of the
correlation between $F_{W}^{208}$ and $F_{W}^{48}$, suggesting that
PREX-II and CREX provide complimentary information. Indeed, whereas
PREX-II places powerful constrains on bulk nuclear-matter properties
(primarily $L$), $^{48}$Ca -- with a significant larger
surface-to-volume ratio than $^{208}$Pb -- may help constrain
better surface properties of nuclear structure models by providing a powerful bridge between
ab-initio calculations and density-functional theory. 

In further comparing models in Fig.\,\ref{Fig1}, we observe that the
linear behavior displayed in the figure is characterized by nearly
equal slopes for all models but different intercepts. It is
interesting to note that there is a significant spread even among the
RMF variants.  Recall that DDME introduces density dependence directly
into the meson-nucleon couplings whereas both NL3 and FSU incorporate
density-dependent effects through non-linear meson self-interactions
and mixed terms.
In particular, we note that models that predict  the same
$r_{\rm skin}^{208}$ show large variations in $F_{W}^{48}$, suggesting that a
measurement of the neutron radius of $^{208}$Pb is 
unable to constrain the
neutron radius in $^{48}$Ca\,\cite{(Pie12)}. This is
likely to suggest significant differences in the surface properties of
the models used. Preliminary explorations along these lines are now
in progress.

\subsection{Covariance analysis}

Having estimated the systematic uncertainties generated by various
models, we now proceed to implement the correlation analysis directly in 
terms of the covariance matrix. To this end, we compute
correlation coefficients (\ref{eq:correlator})
for all the models considered --
 directly in terms of their own covariance matrix (\ref{Mij}).
\begin{figure}
\centerline{\includegraphics[width=0.99\linewidth]{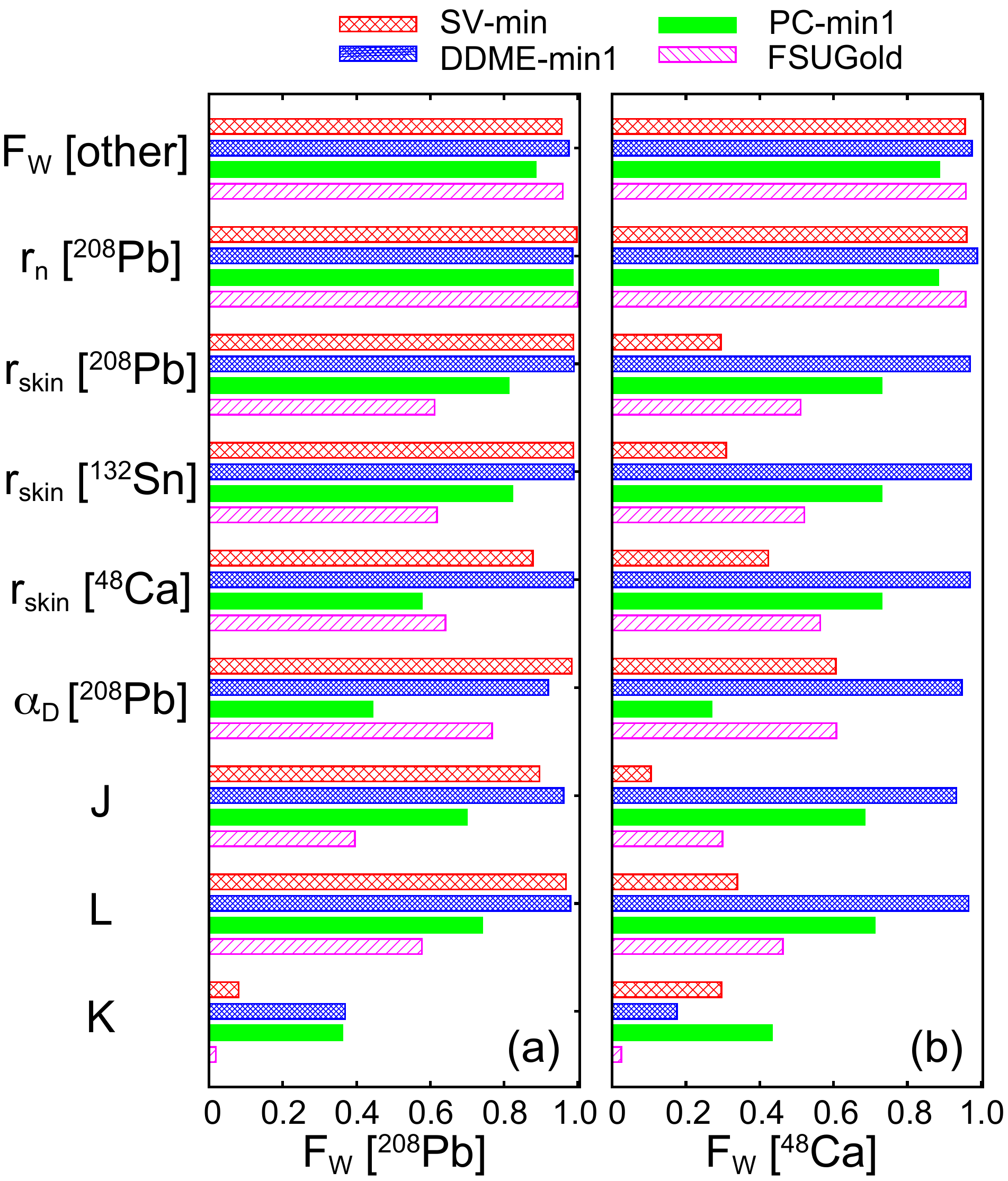}}
\hspace*{1em}
\caption{(Color online) Correlation coefficients (\ref{eq:correlator}) derived from a covariance
analysis between the weak-charge form factors of ${}^{208}$Pb ($F_{W}^{208}$)
and ${}^{48}$Ca ($F_{W}^{48}$), and a variety of nuclear observables, including 
the strong isovector indicators such as neutron skin, electric dipole polarizability, symmetry energy $J$,
slope of the symmetry energy $L$, and a strong isoscalar indicator: incompressibility 
$K$ at saturation density. In all cases correlation coefficients  are obtained 
from the covariance matrix associated with each model.}
\label{Fig2} 
\end{figure}
Figure\,\ref{Fig2}(a) displays correlation coefficients
between $F_{W}^{208}$ and a suitable selection of physical observables
and bulk parameters of infinite nuclear matter. The analogous information 
on $F_{W}^{48}$ is shown in Fig.~\ref{Fig2}(b). The
accurately-calibrated models included in this comparison are SV-min (a
SHF model) and three RMF variants: DDME-min1, PC-min1, and
FSUGold. The first (topmost) entry illustrates the  excellent correlation
 between $F_{W}^{208}$ and $F_{W}^{48}$ within each
model. This is reminiscent of the strong correlation -- within each
model family -- observed in Fig.\,\ref{Fig1}(d). However, recall that as
systematic uncertainties across the various models are assessed, the
correlation weakens significantly. These findings reinforce the argument in favor of
combined measurements of $F_{W}^{208}$ and $F_{W}^{48}$. Indeed, both
PREX-II and CREX will provide invaluable information in discriminating
among various SCMF models.

The second entry confirms the strong correlation
between $r_{n}^{208}$ and  $F_{W}^{208}$. This suggest that although 
$q_{{}_{\rm PREX}}r_{n}^{208}\!\gtrsim\!1$, thereby invalidating a
direct extraction of $r_{n}^{208}$ from $F_{W}^{208}$, measuring
the form factor provides a strong constraint on the neutron 
radius\,\cite{Furnstahl:2001un}. Still, we stress that the cleanest
comparison between theory and experiment is directly in terms 
of the experimentally measured form factor. We will look into this 
connection in more detail in Sec.\,\ref{sec:sensi}.

The correlations with  $r_{\rm skin}$ may seem surprising at
first glance. There is an apparent dichotomy between SV-min and
DDME-min1 on the one hand, and PC-min1 and FSUGold on the
other. Whereas the former display a strong correlation between
$r_\mathrm{skin}^{208}$ and $F_{W}^{208}$, the correlation weakens
significantly for the latter.  The apparent contradiction may have its
origin in the underlying fitting protocols
and different density dependence.  For example, both SV-min
and DDME-min1 include the charge radius of ${}^{208}$Pb ($r_{\rm
  ch}^{208}$) with its very small experimental
error\,\cite{Angeli:2013} in the fit. Such a small error strongly
constrains the linear combination of model parameters sensitive to the
charge radius.  This implies that the exploration of the model
landscape is ``locked'' at the experimental value of $r_{\rm
  ch}^{208}$ (or equivalently $r_{p}^{208}$).  Thus, any changes in
$r_\mathrm{skin}^{208}$ around the optimal model are essentially
generated by the corresponding changes in $r_{n}^{208}$.  The charge
radius of ${}^{208}$Pb was also included in the calibration procedure
of the FSUGold functional\,\cite{Todd-Rutel:2005fa}. However, in
contrast to SV-min and DDME-min1, no covariance matrix was extracted
at that time. Thus, the FSUGold correlation coefficients presented in
this work were obtained from a simplified covariance analysis of Ref.~\cite{(Fat11)} that did
not include $r_{\rm ch}^{208}$ into the fit.  
Clearly,
as we develop next-generation EDFs, their  optimization procedure  should always
generate both the optimal model as well as the covariance matrix. We
note that in the case of the correlation between $F_{W}^{48}$ and
$r_\mathrm{skin}^{48}$, DDME-min1 remains as the sole model displaying
a strong correlation; in the case of SV-min, the correlation is much weaker.
Again, this difference may originate from the
various fitting protocols.

In Ref.\,\cite{(Rei10)} the electric dipole polarizability in
${}^{208}$Pb was identified as a strong isovector indicator that is
strongly correlated to $r_\mathrm{skin}^{208}$. Here too we find a
strong correlation between $F_{W}^{208}$ and
$\alpha_{D}^{208}$, except for  PC-min1 where this
correlation appears to be fairly weak. Moreover, we note that the
correlation between $\alpha_{D}^{208}$ and the weak-charge form
factor of ${}^{48}$Ca weakens significantly, with the exception of
DDME-min1. Within the next few years we expect that CREX and PREX-II
will provide accurate measurements of the neutron radius of
${}^{48}$Ca and ${}^{208}$Pb with anticipated errors of $0.02$\,fm and
$0.06$\,fm, respectively. We note that a high-precision
measurement of $\alpha_{D}^{208}$ is now available\,\cite{(Tam11)} and
that the corresponding measurement on ${}^{48}$Ca is presently being
analyzed~\cite{Birkhan:2013}. 
When combined, these four key isovector
indicators will provide the critical input for the calibration of EDFs of
increasing sophistication.

Finally, correlations between $F_{W}^{208}$ and bulk parameters of
infinite nuclear matter display a large model dependence. For example,
both SV-min and DDME-min1 display a strong correlation between
$F_{W}^{208}$ and the symmetry energy $J$ and the slope of the
symmetry energy $L$ at saturation density.
This appears not to be the
case for PC-min1 and FSUGold.  As mentioned earlier, this reflects the
simplified covariance analysis with FSUGold that failed to include a
tightly constrained charge radius of ${}^{208}$Pb into the
fit. Finally, as an illustration, we show how a strong isoscalar
indicator such as the incompressibility of symmetric
nuclear matter $K$ is weakly correlated with both weak-charge form
factors in all the models.

\subsection{Momentum-transfer sensitivity of the weak-charge form factor}
\label{sec:sensi}

\begin{figure}
\centerline{\includegraphics[width=0.85\linewidth]{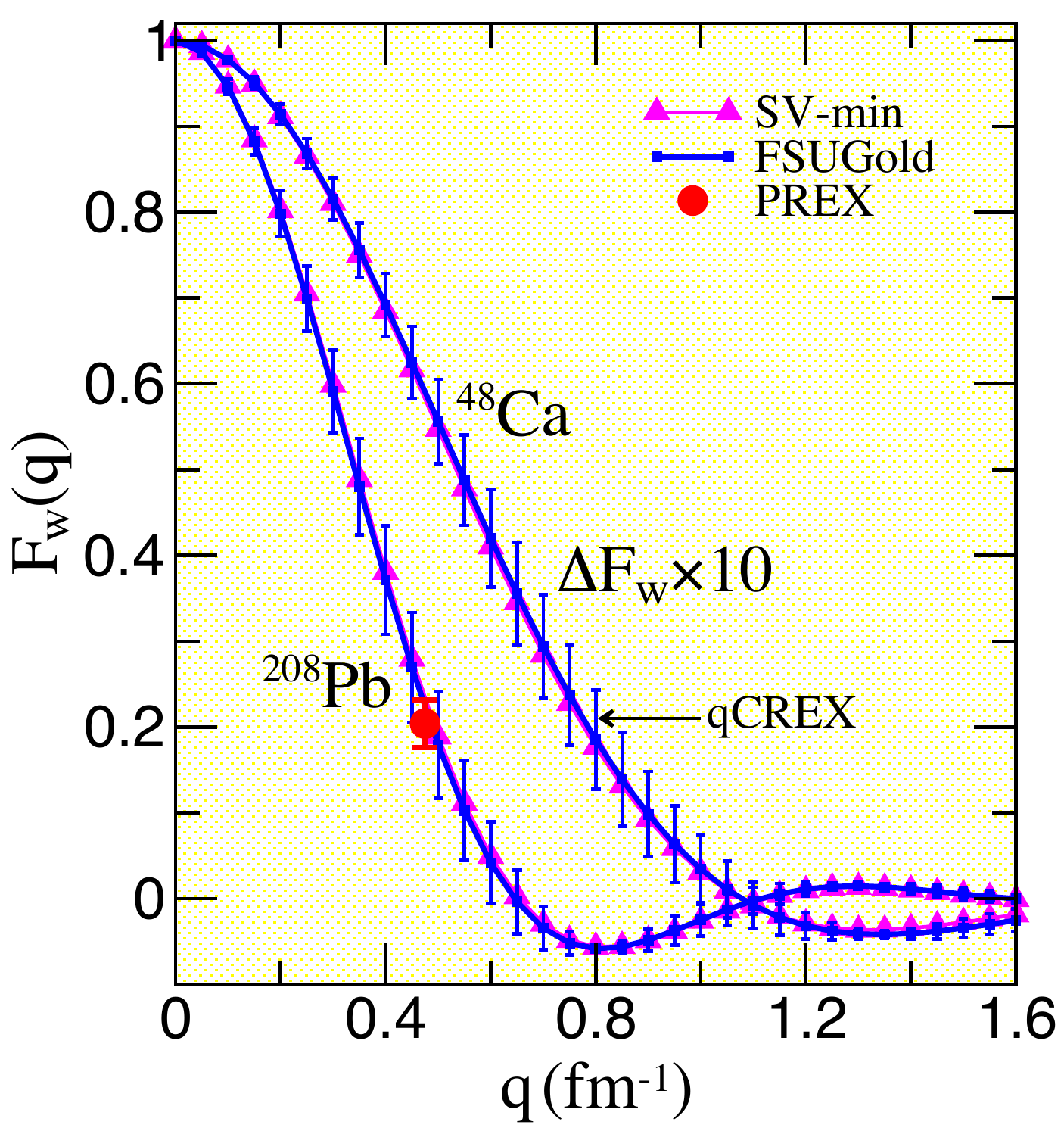}}
\caption{(Color online) Weak-charge  form factors with corresponding
theoretical errors for ${}^{48}$Ca and ${}^{208}$Pb as predicted by 
SV-min and FSUGold. Note that the theoretical error bars have been 
artificially increased  by a factor of 10. Indicated in the figure are the values of the momentum 
transfer appropriate for PREX-II ($q\!=\!0.475\,{\rm fm}^{-1}$) and 
CREX ($q\!=\!0.778\,{\rm fm}^{-1}$).}
\label{Fig3} 
\end{figure}
So far we have only considered the weak-charge form factors at
the relevant momentum transfers of CREX and PREX-II, namely, 
$q_{{}_{\rm CREX}}\!=\!0.778$\,fm$^{-1}$ and
$q_{{}_{\rm PREX}}\!=\!0.475$\,fm$^{-1}$. We now explore the
sensitivity of the correlation between the neutron radius and the
weak-charge form factor as a function of the momentum transfer 
$q$. This is particularly relevant because the optimal momentum 
transfer emerges from a compromise between the elastic cross 
section -- which falls rapidly with $q$ -- and the parity-violating 
asymmetry -- which is proportional to $q^{2}$. However, it is 
{\it a-priori} unclear whether at an optimal momentum 
transfer (which is not small) the correlation between the weak-charge 
form factor and the neutron radius is strong. 

We begin by displaying in Fig.\,\ref{Fig3} the weak-charge form factor
for both ${}^{48}$Ca and ${}^{208}$Pb with their associated
theoretical uncertainties as a function of $q$. Note that in order to
make the theoretical errors visible they had to be amplified by a
factor of 10. The results show clearly the faster falloff of $F_{W}^{208}(q)$ due to its larger weak-charge radius.  In
particular, this allows CREX to go to a higher momentum transfer where
the parity-violating asymmetry is larger. At the  values  of the proposed momentum
transfers, the predicted form factors are almost equal, i.e., 
$F_{W}^{48}\!\approx\!F_{W}^{208}\!\approx\!0.2$. Note that the
predictions from the non-relativistic SV-min and the relativistic
FSUGold agree very well with each other and both are consistent with the PREX
measurement. We emphasize that although some model-dependent
assumptions must be invoked in extracting the neutron radius from a
measurement of the form factor, such assumptions are ultimately
unnecessary. This is because one can always compare the theoretical
form factors directly with  experiment.

\begin{figure}
\centerline{\includegraphics[width=0.95\linewidth]{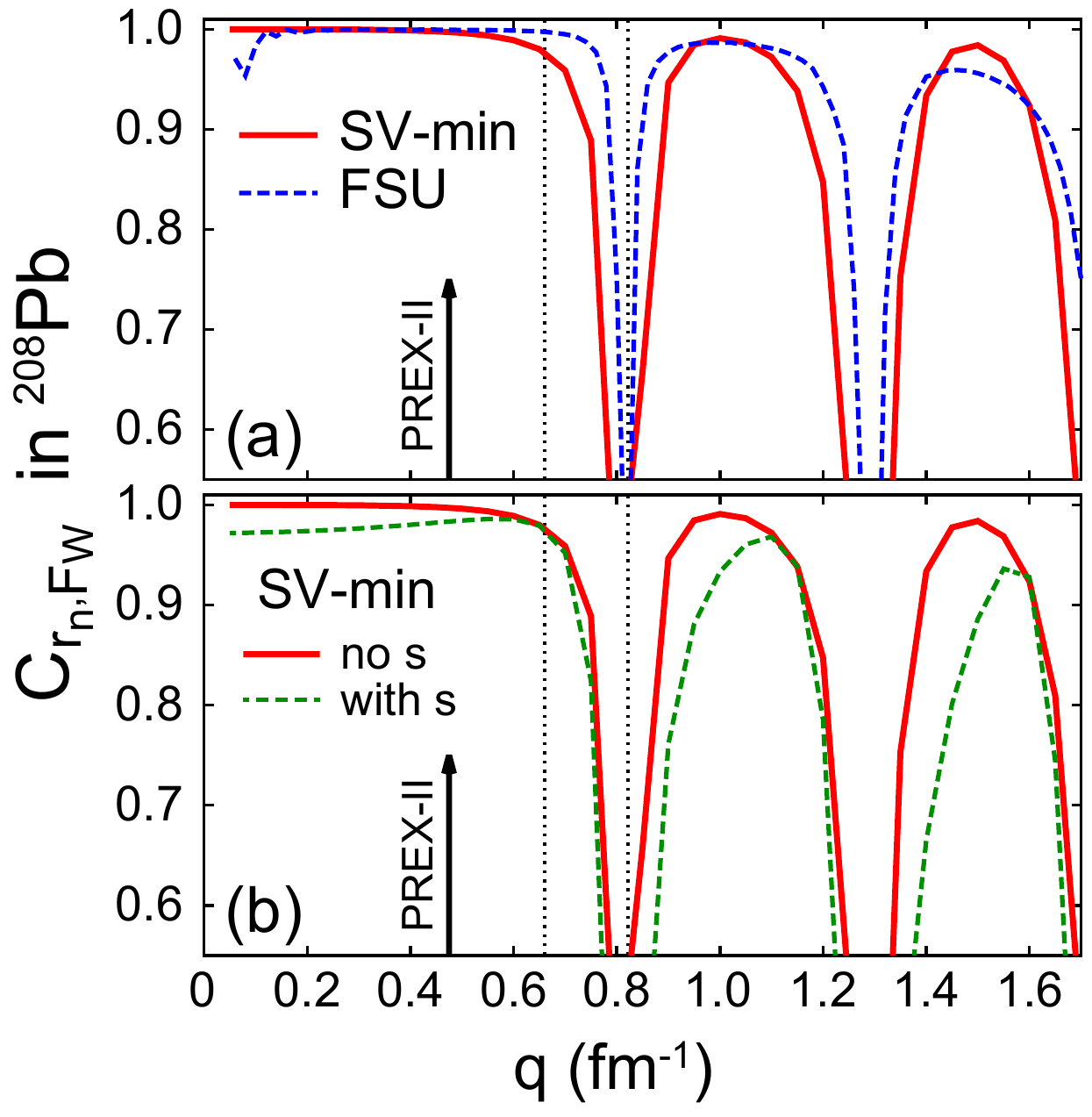}}
\caption{(Color online) Correlation coefficient (\ref{eq:correlator})  between $r_{n}^{208}$ and $F_{W}^{208}(q)$
as a function of the
  momentum transfer $q$. Panel (a) shows the absolute value of the
  correlation coefficient predicted by SV-min and FSUGold assuming no
  strange-quark contribution to the nucleon form factor. Panel (b)
  shows the impact of including the experimental uncertainty in the
  strange-quark contribution to the nucleon form factor. The arrow marks  the PREX-II momentum transfer of $q\!=\!0.475\,{\rm
    fm}^{-1}$. The first dashed vertical line indicates the position
  of the first zero of $F_W^{208}(q)$, the second one marks the position of
  the first maximum of $\left|F_W^{208}(q)\right|$ (from which the surface
  thickness can be deduced).}
\label{Fig4}
\end{figure}
We now explore the correlation between $r_{n}^{208}$ and $F_W^{208}(q)$ for a range of momentum transfer. In 
Fig.\,\ref{Fig4}(a)  we compare the (absolute value) of the correlation as
predicted by SV-min and FSUGold. At small momentum transfer, the form
factor behaves as
$F_{W}(q)\!\approx\!1\!-\!q^{2}r_{W}^{2}/6\!\approx\!1\!-\!q^{2}r_{n}^{2}/6$
so the correlation coefficient is nearly one. Note that we have used
the fact that the weak-charge radius $r_{W}$ is approximately equal to
$r_{n}$ \cite{(Hor01)}. Also note that although at the
momentum transfer of the PREX experiment the low-$q$
expression is not valid, the strong correlation is still
maintained. Indeed, the robust correlation is maintained at all $q$-values, 
except for diffraction minima and maxima. Given the similar patterns 
predicted by SV-min and
FSUGold, we suggest that the observed $q$-dependence of the
correlation with $r_n$ represents a generic model feature.

Figure\,\ref{Fig4}(b) displays the same correlation, but now
we also include the experimental uncertainty on the strange-quark form
factor. Although the strange-quark contribution to the electric form
factor of the nucleon appears to be very small\,\cite{Liu07a}, there
is an experimental error attached to it that we want to explore. For
simplicity, only results using SV-min are shown with and
without incorporating the experimental uncertainty on s-quark.
We note that an almost perfect correlation
at low-to-moderate momentum transfer gets diluted by about 6\% as the
uncertainty in the strange-quark contribution is included. Most interestingly,  the
difference almost disappears near the actual PREX point, lending confidence that the
experimental conditions are ideal for the extraction of $r_{n}^{208}$. Finally, given that the strong correlation
between the neutron radius and the form factor is maintained up to the
first diffraction minima (about $q\!=\!1.2\,{\rm fm}^{-1}$ in the case
of ${}^{48}$Ca) the CREX experimental point lies safely within this
range (figure not shown).

\section{Conclusions and Outlook}
\label{sec:conclusions}

In this survey, we have studied the potential impact of the proposed  PREX-II and
CREX measurements on constraining the isovector sector of the nuclear EDF. In
particular, we explored correlations between the weak-charge form
factor of both ${}^{48}$Ca and ${}^{208}$Pb, and a variety of
observables sensitive to the symmetry
energy. We wish to emphasize that we have chosen the weak-charge form factor rather
than other derived quantities -- such as the weak-charge (or neutron)
radius -- since $F_W$ is directly accessed by
experiment. To assess correlations among observables, two different
approaches have been  implemented. In both cases we relied exclusively on
models that were accurately calibrated to a variety of ground-state
data on finite nuclei. In the ``trend analysis'', the parameters of the optimal
model were adjusted in order to systematically change  the
symmetry energy, and the resulting impact on nuclear observables was monitored. In the  ``covariance
analysis'', we obtained correlation coefficients by relying exclusively
on the covariance (or error) matrix that was obtained in the process of model optimization.

We verified that the neutron skin of
${}^{208}$Pb provides a fundamental link to the equation of state of
neutron-rich matter.
The landmark PREX experiment achieved a very small systematic error on
$r_{n}^{208}$ that
suggests that reaching the  total error of $\pm 0.06$\,fm anticipated in PREX-II
is realistic.  We also concluded that an
accurate determination of $r_{\rm skin}^{208}$ is
insufficient to constrain the  neutron skin of
${}^{48}$Ca. Indeed, because of the significant difference in the
surface-to-volume ratio of these two nuclei, there is a considerable
spread in the predictions of the models\,\cite{(Pie12)}. Given that
CREX intends to measure $r_{\rm skin}^{48}$ with an unprecedented error
of $\pm 0.02$\,fm, this model dependence can be tested
experimentally \cite{(Kor13a)}. In addition, as discussed in Sec.~\ref{intro}, there are several advantages for nuclear theory  in measuring
the neutron radius of both ${}^{48}$Ca and ${}^{208}$Pb.  We have
verified that at the momentum transfer selected for PREX-II
(0.475\,fm${}^{-1}$) there is a large sensitivity of the weak-charge
form factor to the neutron radius of ${}^{208}$Pb; a similar
conclusion was obtained in the case of CREX. Finally, we estimated the contribution from the strange-quark uncertainty 
on the electric form factor error budget. We concluded that this contribution is very small  near the actual PREX $q$-value.

In summary, although PREX-II provides a powerful constraint on the
slope of the symmetry energy $L$, the neutron radius of ${}^{48}$Ca is
sensitive to nuclear dynamics that goes well beyond $L$. Thus, CREX in
combination with PREX-II will constrain different aspects of the nuclear
EDF. Moreover, we have reconfirmed that the electric-dipole
polarizability in ${}^{208}$Pb represents a strong isovector
indicator.  Hence, we strongly advocate measurements of the neutron
radius and electric-dipole polarizability in ${}^{48}$Ca. Together,
these four observables -- neutron radii and dipole polarizabilities in
both ${}^{48}$Ca and ${}^{208}$Pb --  will form a critical set of isovector
indicators that will provide essential  constraints on 
nuclear density functionals of the next-generation.

\begin{acknowledgments}
Useful discussions with the CREX collaboration are gratefully
acknowledged. This work was supported  by the U.S. Department of
Energy (DOE) under Contracts No.
DE-FG05-92ER40750 (FSU), No. DE-FG02-96ER40963 (UTK), and by the BMBF
under Contract No. 06ER9063.
\end{acknowledgments}

\appendix
\section{The weak-charge form factor}
\label{sec:FW}

Here, we briefly summarize the computation of the weak-charge form factor as
detailed in Ref.~\cite{(Hor01)}.  The basic input are the local proton and
neutron density distributions, $\rho_p$ and $\rho_n$, respectively. Accounting for
magnetic contributions would require also the spin-orbit current (for
SHF) or the tensor current (for RMF) \cite{(Hor12)}. We ignore these as they add only
a small correction, which is not important for this survey.
The proton and neutron densities are normalized in the usual way: $\int d^3r\rho_p=Z$
and $\int d^3r\rho_n=N$. Note that Ref. \cite{(Hor01)} uses the
invariant four momentum $Q^2$ and the spatial momentum $q$ side by
side. They are related by $q=\sqrt{Q^2}$. We use only $q$
throughout.

We assume spherically symmetric systems,
i.e.,  $\rho(\mathbf{r})=\rho(r)$ where $r=|\mathbf{r}|$.
In general,  $F(q)$ and $\rho(r)$ are
connected through the Fourier transformation \cite{Fri82a}
\begin{subequations}
\begin{eqnarray}
  F(q)
  &=&
  \int d^3r\,e^{\mathrm{i}\mathrm{q}\cdot\mathrm{r}}\rho(r)
\nonumber\\
  &=&
  4\pi\int_0^\infty dr\,r^2\,j_0(qr)\rho(r),
\label{eq:formf-rho}\\
  \rho(r)
  &=&
  \int\frac{d^3q}{8\pi^3}\,e^{-\mathrm{i}\mathrm{q}\cdot\mathrm{r}}F(q)
\nonumber\\
  &=&
  \frac{1}{2\pi^2}\int_0^\infty dq\,q^2\,j_0(qr)F(q).
\end{eqnarray}
\end{subequations}
The transformation applies to any local density, for 
protons $\rho_p\longleftrightarrow F_p$,
neutrons $\rho_n\longleftrightarrow F_n$,
and the weak-charge density $\rho_W\longleftrightarrow F_W$.

We prefer to formulate the weak-charge distributions in terms of the
form factor because the necessary folding operations become much simpler in
the Fourier space.
The weak charge form factor normalized to one at $q=0$ can be written as:
\label{eq:formfweeak}
\begin{equation}
  F_W^{\mbox{}}(q)
  =
  e^{a_\mathrm{cm}q^2}
  \frac{G_n^Z(q)F_n(q)+G_p^Z(q)F_p(q)}
       {G_n^Z(0)F_n(0)+G_p^Z(0)F_p(0)},
\label{eq:FW}
\end{equation}
with
\begin{subequations}
\begin{eqnarray}
  G_p^Z
  &=&
  \mathcal{N}_p\left[
  \frac{G_p-G_n}{4}-S_{\Theta_W}G_p-\frac{G_s}{4}
  \right],
\\
  G_n^Z
  &=&
  \mathcal{N}_n\left[
  \frac{G_n-G_p}{4}-S_{\Theta_W}G_n-\frac{G_s}{4}
  \right],
\\
  G_s(q)
  &=&
  \rho_s\frac{\hbar^2q^2/(4c^2M^2)}{1+4.97\,\hbar^2q^2/(4c^2M^2)},
\label{eq:Gs}
\\
  \mathcal{N}_p  
  &=&
  \frac{0.0721}{1-4\sin^2(\Theta_W)}
  \;,\;
  \mathcal{N}_n
  =
  0.9878,
\label{eq:renorm-rad}
\end{eqnarray}
\end{subequations}
where $G_p$ and $G_n$ are  the standard proton and neutron
electro-magnetic form factors, respectively; $G_s$ is the strange-quark
electric form factor;  $S_{\Theta_W}=
  \sin^2(\Theta_W) = 0.23$; $\rho_s= (-0.24\pm 0.70)$\,fm;
$M$ is the average nucleon mass; and $a_\mathrm{cm}$ a parameter
for the center-of-mass  (c.m.) correction. The renormalization factors $\mathcal{N}_p$,
$\mathcal{N}_n$ take into account the
radiative corrections to the weak charge \cite{PDG}.
They guarantee that the weak-charge becomes 0.0721 for the proton
and $-0.9878$ for the neutron.
The simple renormalization by a constant factor assumes that the
corrections do not change significantly over the range of $q$ 
relevant for the PREX measurements.

The strength $\rho_s$ of the $s$-quark coupling and its
uncertainties are taken from Refs.~\cite{HAPPEX07,Liu07a}. These two
evaluations agree in the strength and have slightly different values for
the uncertainties. In this work, we took the average of both. 

\begin{table}[b]
\begin{center}
\begin{tabular}{crrrrrrrr} 
\\
\hline\noalign{\smallskip} 
  & $a_1$ & $a_2$ & $a_3$ & $a_4$ & $b_1$ & $b_2$ & $b_3$ & $b_4$\\ 
\noalign{\smallskip}\hline\noalign{\smallskip}
 $E_{0}$ & 2.2907 & -0.6777 & -0.7923 & 0.1793 &
           15.75 & 26.68 & 41.04 & 134.2            \\
 $E_{1}$ & 0.3681 & 1.2263 & -0.6316 & 0.0372 &
            5.00 & 15.02 & 44.08 & 154.2 \\
\noalign{\smallskip}\hline
\end{tabular}
\end{center}
\caption{\label{tab:iform}
Parameters of the model (\ref{eq:FPol}) for the nucleon form factors.
 The constants $b_i$ are given in units fm$^{-2}$.
$E_0$ is the isoscalar ($I$=0) electric form factor and $E_1$
the isovector one ($I$=1).
The 
form factors are taken from \cite{Wal86aPC}.
}
\end{table}

\begin{subequations}
A word is in order about the c.m. correction. 
The variance of the c.m. momentum
$\langle\hat{P}_{c.m.}^2\rangle$ is computed in SCMF models  from the actual
wave function to define the coefficient
\begin{equation}
  a_\mathrm{cm}
  =
  \frac{\hbar^2}{8\langle\hat{P}_{c.m.}^2\rangle}.
\end{equation}
One often uses a simple estimate for the c.m. correction
energy from a harmonic oscillator shell model. In this context, it is
consistent to make the replacement:
\begin{equation}
  a_\mathrm{cm}
  =
  \frac{1.58}{6.0A^{2/3}}\,{\rm fm}^2,
\end{equation}
where $A$ is the mass number.
\end{subequations}

The weak-charge form factor is expressed in terms of the intrinsic nucleon form factors.
We use here the traditional form of Simon \& Walther
\cite{Sim80aE,Wal86aPC}.  It parametrizes isoscalar and isovector
form factors as a sum of dipole terms:
\begin{equation}
  G_\mathrm{typ}^\mathrm{(S)}(q)
  =   
  \sum_{\nu=1}^4 \frac{a_{\mathrm{typ},\nu}}{1+q^2/b_{\mathrm{typ},\nu}}
\label{eq:FPol}
\end{equation}
with ${\rm typ}\in\{\mbox{``$E$, $I$=0'', ``$E$, $I$=1''}\}$
and with parameters  listed in Table \ref{tab:iform}.
The proton and neutron Sachs form factors are:
\begin{subequations}
\label{eq:formfS-intr}
\begin{eqnarray}
  G_{E,p}^\mathrm{(S)}
  &=& 
  \frac{1}{2} ( G_{E,I=0}^\mathrm{(S)} + G_{E,I=1}^\mathrm{(S)} ),
\label{eq:GEpS}
\\
  G_{E,n}^\mathrm{(S)}
  &=& 
  \frac{1}{2} ( G_{E,I=0}^\mathrm{(S)} - G_{E,I=1}^\mathrm{(S)} ),
\label{eq:GEnS}
\end{eqnarray}
\end{subequations}

\bibliographystyle{apsrev4-1}
\bibliography{weak-charge}

\end{document}